\documentstyle[aps,prb,preprint]{revtex}

\begin{document}

\draft

\title{\bf The High Magnetic Field Phase Diagram of a Quasi-One
Dimensional Metal}

\author{J.S. Qualls,$^1$, C.H. Mielke,$^2$, J.S. Brooks,$^1$ L.K.
Montgomery,$^3$ D.G. Rickel,$^2$ N. Harrison,$^2$ and S.Y. Han $^1$}

\address{  $^1$NHMFL, Florida State University, Tallahassee,
Florida 32310 USA \\$^2$Los Alamos National Laboratory/NHMFL, MS E536, Los
Alamos,New Mexico 87545 USA \\ $^3$Department of Chemistry, Indiana
University,
Bloomington, IN 47405 USA}
\date{\today}
\maketitle
\begin{abstract}

We present a unique high magnetic field phase of the quasi-one dimensional organic
conductor (TMTSF)$_2$ClO$_4$. This phase, termed "Q-ClO$_4$", is obtained
by rapid thermal quenching to avoid ordering of the ClO$_4$ anion. The
magnetic field dependent phase of Q-ClO$_4$ is distinctly different from
that in the extensively studied annealed material. Q-ClO$_4$ exhibits a
spin density wave (SDW) transition at $\approx$ 5  K which is strongly
magnetic field dependent. This dependence is well described by the
theoretical treatment of Bjelis and Maki. We show that Q-ClO$_4$ provides
a new B-T phase diagram in the hierarchy of low-dimensional organic metals (one-dimensional towards two-dimensional), and describe the temperature dependence of the of the quantum oscillations observed in the SDW phase.

\end{abstract}
\newpage

\vskip 5mm

\narrowtext

Quasi-one dimensional organic metals have the general character of a large
bandwidth along the molecular stacking (chain) direction, followed by
significantly smaller band widths in the inter-chain and inter-plane
directions\cite{Ish}. For the Bechgaard salts these ratios are ($t_a$, $t_b$, $t_c$:
1000, 250, 3 meV) respectively\cite{Bec}. For sufficiently
small transverse bandwidths, a one-dimensional conductor will undergo an
instability, at a critical temperature, to an insulating ground state.
Following Yamaji\cite{Yam}, this temperature has an anisotropic bandwidth
dependence in terms of the so called "imperfect nesting parameter",
$\varepsilon_0$ = $t_b^2/t_a$. Hence the more two-dimensional the material is, the larger
$\varepsilon_0$ will  be, and the lower the temperature where the
instability, or "nesting" will occur. In the case of the Bechgaard  salts, a spin density wave (SDW) ground state is formed.  For sufficiently large
$\varepsilon_0$, the low temperature ground state remains metallic, but
high magnetic fields can effectively reduce $\varepsilon_0$ (i.e. drive the
system more one-dimensional), and a field induced spin density wave state
(FISDW) can be stabilized\cite{Gor}. This latter phenomena has been the
subject of extensive experimental and theoretical work\cite{Ish}.

In this communication we consider a system where $\varepsilon_0$ is large,
but where an SDW state still forms at a temperature T$_{SDW}$ = 5 K. This allows a 
unique situation where T$_{SDW}$ increases by a factor of two in high magnetic fields due to the close competition between $\varepsilon_0$ and the magnetic energy.
To accomplish this we employ the well-studied
organic conductor (TMTSF)$_2$ClO$_4$.\cite{Ish} What is particular to our
approach is that the material has been prepared in a very rapidly (30K/sec)
thermally quenched state (i.e. Q-ClO$_4$) to preserve the high
temperature electronic structure, which is comprised of two, open orbit,
warped Fermi surface sheets. Otherwise, if the material is slowly cooled,
the tetrahedral anion ClO$_4$ undergoes an ordering transition
around 24 K, the unit cell doubles in the b direction, and the resulting
Fermi surface becomes more complex\cite{Mck}. Generally, it is this relaxed state of the material (hereafter R-ClO$_4$), which has been
most extensively studied. In contrast, Q-ClO$_4$
is in the class of Bechgaard salts\cite{Bec} (TMTSF)$_2$X where X = PF$_6$,
AsF$_6$, NO$_3$ which form an SDW state at ambient pressure below a
transition temperature $T_{SDW}$ (12 K for X = PF$_6$ and AsF$_6$, 10K for
X = NO$_3$, and 5 K for X = Q-ClO$_4$).  In these cases
a high magnetic field improves the nesting condition\cite{Aud,Uji,Bro}, and
$T_{SDW}$ increases with magnetic field. We have determined 
that not only does
this happen for the case of Q-ClO$_4$ , but the effect is the most
dramatic, and leads to a new phase diagram in the hierarchy of nesting parameters ( 7 K $\le$ $\varepsilon_0$ $\le$ 22.5 K ) in the Bechgaard salts. 

The measurements reported here were carried out on two samples in 50 T
(sample $\#1$) and 60 T (sample $\#2$) pulsed-field magnets at the Los
Alamos National Laboratory. Electrical transport contacts were made via
graphite paint and 12$\mu$$m$ gold wires, with a dc four-terminal technique
with a current of 50 $\mu$$A$. The current was applied transverse to the
layers along the c-axis, as were the voltage contacts. The magnetic field is also along the c-axis. To ensure that the
samples fully quenched, the samples were put in direct contact with liquid
helium from room temperature as rapidly as possible. Estimated cooling
rates were of order 30 K/sec or greater.  The use of graphite paint appeared to greatly enhance the reliability of the contacts and to reduce degradation (cracking) of the samples during the rapid cool-downs. Systematic temperature
measurements for each run were performed with a single quench to preserve
the anion disorder in the samples. 

In Fig. 1a we show a summary of the magnetoresistance measurements for
sample $\#1$ above 5 K, and in Fig. 1b similar measurements at lower temperatures for both samples $\#1$ and $\#2$ are presented.  In the inset of Fig. 1a  the temperature dependence of the c-axis resistivity is shown at zero magnetic field. Here the upturn is
the onset of the spin density wave transition, as established by, for
instance NMR studies \cite{Del}. In Fig. 2 we show an expanded view of the
behavior of the magnetic field
dependent resistance, where the data has been plotted vs.
the square of the magnetic field. All data shown is for temperatures higher than $T_{SDW}$
(B=0). Here the magnetoresistance remains small, and quadratic in field, until a critical field (hereafter $B_{SDW}$) is reached. We
have fit the 10.6 K resistance data in Fig. 2 to the quadratic function (expected for the case of an
open orbit metal) below the point where the slope changes. Aside from a
weak temperature
dependence, all of the data follow this general functional dependence until
the point $B_{SDW}$ is reached.
$B_{SDW}$ is temperature dependent, and is manifested as a change in the
field dependence of the magnetoresistance. This defines the threshold field
between the metallic and SDW ground
states. Above
$B_{SDW}$, additional structure appears (hereafter
$B'_{SDW}$) where the field dependence of the magnetoresistance changes
again. At higher fields quantum oscillations of frequency F=190 T
become evident. We will return to these last two points in the discussion below.

In Fig. 3 we summarize the dependence of $B_{SDW}$ and $B*$ in terms
of the corresponding temperatures $T_{SDW}$ and $T*$ based on the
analysis of Fig. 2 and the zero field value from  the inset
of Fig. 1a.
The new phase diagram, as defined by Fig. 3, is the main result of the
present work. To put the new data for Q-ClO$_4$ in perspective, we have
included previous results for AsF$_6$, PF$_6$, and NO$_3$, (Refs.
\cite{Uji,Aud,Nau}) for the field dependence of $T_{SDW}$ in terms of the
theoretical framework, which describes the magnetic field dependence of
$T_{SDW}$ as given by Bjelis and Maki in the form\cite{Bje}

\begin{equation}
\ln{\left[ {T_{SDW} \over T_{SDW0}} \right]}
~\cong~ \sum_{l=-\infty}^{\infty} J^2_l (e_0)
\left\{ Re\Psi({1 \over 2}+2ilx_1)-\Psi({1 \over 2}) \right\}.
\end{equation}

\noindent Here $e_0$=$\varepsilon_0$/$\omega_b$, where $\varepsilon_0$
is the imperfect nesting parameter described above, and
$\omega_b$=$ev_FbB$.                                        The latter is
the effective cyclotron frequency along the b-axis, where  $v_F$ is   the
Fermi velocity.
$J_l$ and $\Psi$ are           Bessel and digamma functions respectively,
and $x$$_1$ = $\omega$$_b$/4$\pi$$T_{SDW}$. $T_{SDW0}$     corresponds to
the transition temperature for perfect nesting. This expression, in an
asymptotic form\cite{Aud}, successfully describes the field dependence of
$T_{SDW}$ for (TMTSF)$_2$NO$_3$  Ref. \cite{Aud} and (TMTSF)$_2$PF$_6$
Ref.\cite{Bro}. For Q-ClO$_4$, Eq. 1 is also applied, but its full
(non-asymptotic) form must be used to obtain proper convergence at low
magnetic fields due to the large imperfect nesting parameter $\varepsilon_0$ needed to describe the data. (The fitting parameters for all three materials are listed
in the caption of Fig. 3.) It is clear that $T_{SDW}$(B =0) decreases with
increasing $\varepsilon_0$. However, since the general Fermi surface
topology is very similar, all three
materials approach a common transition temperature in the high field limit.

For completeness, the phase diagram of R-ClO$_4$ is also shown\cite{Mck} in Fig. 3. A point that must be clearly made is that the Q-ClO$_4$ phase
diagram is very different from that studied in the R-ClO$_4$
case. For R-ClO$_4$ the Fermi surface topology involves double open orbit
sheets due to the anion ordering, the imperfect nesting is so large that at
zero field the system is metallic, even superconducting (T$_c$ = 1.2 K), and the maximum
field induced $T_{SDW}$ is less than 6 K. Another factor which distinguishes Q-ClO$_4$ from R-ClO$_4$ is that the quantum
oscillation frequency in the magnetoresistance is 190 T, as confirmed by
this (see below) and previous independent studies\cite{Bro}. In contrast, for R-ClO$_4$
the frequency is 250 T. This confirms that the Fermi surface topologies
are fundamentally different. Returning for a moment to the observation of the $B'_{SDW}$ feature (Figs. 2 and 3),  it is not clear what assignment to make to it (i.e. a SDW sub-phase for instance). It falls outside the prediction of Eq. 1 since there only one transition  ($B_{SDW}$) is predicted. Hence further experimental and theoretical work will be needed to fully understand the significance of $B'_{SDW}$. 

	We next turn to the nature of the oscillations in the magnetoresistance
which appear above $B_{SDW}$ (see Fig. 1). The oscillatory component of the magnetoresistance, plotted as $\Delta R/R$ vs. inverse magnetic field is shown in Fig. 4a and 4b for representative temperatures. (Here R is the non-oscillatory background magnetoresistance.) The oscillation amplitude increases until about 5 K, but then rapidly vanishes at lower temperatures. In Fig. 4c we show
the amplitude of the oscillations (normalized by R )
as a function of temperature. Such oscillations, commonly called "rapid
oscillations" or "RO" appear in a number of the quasi-one dimensional organic
salts, including the Bechgaard salts\cite{Ish} and the
(DMET-TSeF)$_2$AuCl$_2$ salts\cite{Bis}. These oscillations are periodic in
inverse field, with a frequency of between 190 and 250 T. If considered as
closed orbits, they represent about 3\% of the first Brillouin zone. A
general observation is that they only occur when the original open orbit
Fermi surface has undergone a reconstruction. This can happen by an anion
ordering transition\cite{Uji2} and/or nesting of the Fermi
surface\cite{Bro}. In many cases where a spin density wave ground
state is stabilized, the amplitude of the rapid oscillations first
increases with decreasing temperature, but below a characteristic
temperature $T^*$ (typically between 2 to 4K), the amplitude attenuates
exponentially for $T\rightarrow0$. These oscillations have been described
as a magnetic breakdown phenomena\cite{Bro} in an imperfectly nested Fermi
surface topology. However, below $T^*$ there is an improvement of the
nesting condition, such that more of the reconstructed Fermi surface
becomes gapped, and the magnetic breakdown becomes less probable. As shown
in Fig. 4c, and following Ref.\cite{Bro}, this behavior may be modeled by
the standard Lifshitz - Kosevich  description for quantum oscillations
above $T^*$, and an
exponential attenuation below $T^*$. Here the parameters of the model
involve an effective mass m* = 1.2 m$_0$, a Dingle temperature $T_D$ = 7 K,
a $T^*$ = 3.75 K, and a low temperature magnetic breakdown gap of 10 K.

	The Dingle temperature associated with the quantum oscillations discussed above gives insight into the nature of carrier scattering, and therefore disorder, in Q-ClO$_4$. Typically for "clean" organic metals where strong quantum oscillations are observed\cite{Woz}, $T_D$ is of order 1 K.  For Q-ClO$_4$ it is considerably higher ( 7 K).  Since anion disorder is necessary to stabilize the Q-ClO$_4$ state, this could be a possible source for the enhanced scattering. However, similar studies of the quantum oscillations in AsF$_6$ and PF$_6$ anion systems\cite{Bro}, where there is no disorder involved, show a similar value of $T_D$.  Furthermore, given that the SDW state, which should be sensitive to disorder, readily develops in Q-ClO$_4$, we speculate that in spite of the anion disorder, the material is in some sense still relatively clean, and the large value of $T_D$ may be characteristic of the SDW (antiferromagnetic ) nature of these systems.

One other feature of the low temperature data is shown in Fig. 1b. Below
about 3 K there are two distinct changes (shown by the arrows) in the
magnetoresistance, one at around 12 T, and the other around 30 T. At
present we have no explanation for these features. One possibility is that
this behavior is related to the $T^*$ mechanism in some manner. Another is
that these changes are some vestige of the anion-ordered state that only shows
up when the resistivity of the predominantly quenched state is sufficiently
high. (The characteristic fields are close to some of the major FISDW phase
boundaries in the R-ClO$_4$ compound\cite{Kan}.)\\

In summary, the present work shows the evolution of the metal-to-spin
density wave transition for open orbit, quasi-one dimensional metals with
changes in their nesting parameters. By mapping the magnetic field
dependence of the spin-density wave transition, we provide a description of
a new field dependent ground state of the material (TMTSF)$_2$ClO$_4$ where
its low temperature anion ordering transition has been suppressed. The
theory of Bjelis and Maki provide an excellent framework to describe the
general field dependent features of these systems. New
aspects of the Q-ClO$_4$ ground state include the full description of the temperature dependence of the
quantum oscillations in the spin density wave phase, which attenuate exponentially
below 5  K, and a nearly quadratic field dependence of the
resistance in the lower field, high temperature metallic phase of the
system. Some evidence is also present for additional sub-phase structure in the B-T phase diagram, but further work will be needed to make accurate assignments to these features. This work should serve as a guide to future investigations of the magnetic field dependent
mechanisms and properties of quasi-one dimensional metals.

\vskip 0.3cm

{\it Acknowledgments:} We are indebted to K. Maki for testing the
convergence of Eq. 1 at low fields, and to D. Agterberg who gave us
valuable advice on the computation involved. Support from NSF-DMR 95-10427
and 99-71474 (J.S.B.) and a cooperative agreement between NSF-DMR95-27035
and the State of Florida (NHMFL) is acknowledged.

\newpage

\centerline {\bf FIGURE  CAPTIONS}

\vspace{0.2in}

\noindent {\bf Figure 1.}   a) Magnetoresistance $\Delta$R/R(B=0) of thermally
quenched (TMTSF)$_2$ClO$_4$ vs. pulsed magnetic field above 5 K 
(sample \#1). Inset: zero field c-axis resistivity vs. temperature. b) Low
temperature magnetoresistance vs. pulsed magnetic field for sample $\#1$
and $\#2$. Arrows indicate the position of structure in the lowest temperature
data (see text).

\vspace{0.2in}

\noindent {\bf Figure 2.}  Detail of the resistance of sample $\#1$ vs. the
square of the magnetic field for different temperatures. The solid line is
a fit of the high temperature (10.4 K) resistance to a quadratic field
dependence. $B_{SDW}$
($T_{SDW}$) is defined as the point of deviation of the resistance from a
quadratic behavior in field (open circles), and $B'_{SDW}$ ($T'_{SDW}$) is a
second change in the field dependence (see arrows) of the resistance at
higher fields. (Complete temperature labels, left to right, are 5.5, 6.45,
6.5, 6.54, 7.25, 7.52, 7.90, 8.5, 8.98,9.0,10.0, and 10.6 K.)

\vspace{0.2in}

\noindent {\bf Figure 3.}  Temperature vs. magnetic field phase diagram of
Q-ClO$_4$ based on the values of  $T_{SDW}$ vs. $B_{SDW}$ ( also $T'_{SDW}$
vs. $B'_{SDW}$) obtained from Fig. 2. The solid lines are fits of the theoretical
expression of Bjelis and Maki from Eq.1 with the parameters
$v_F$ , $T_0$,  and $\varepsilon_0$ for X = AsF$_6$ ($v_F$ = 2.4 x 10$^5$
$m/s$; $\varepsilon_0$ = 7 K;  $T_0$ = 11.5 K), X = NO$_3$ ($v_F$ = 2.4 x
10$^5$ m/s; $\varepsilon_0$ = 13 K;  $T_0$ = 11.0 K), and X = Q-ClO$_4$ ($v_F$
= 1.1 x 10$^5$ m/s; $\varepsilon_0$ = 22.5 K;  $T_0$ = 13.0 K). The dashed
line is a polynomial fit to $T'_{SDW}$ vs. $B'_{SDW}$. The lower phase diagram is for R-ClO$_4$ after Ref. \cite{Mck}. The finely dashed line is the main second order phase boundary between the metallic and FISDW phases. The other two phase lines T$_H$ and T$_R$ delineate sub-phases (See Ref. \cite{Mck}.)

\vspace{0.2in}

\noindent {\bf Figure 4.}   Temperature dependence of the quantum oscillations (normalized by the background magnetoresistance). The frequency is 190 $\pm$ 5 T. a) Low temperature behavior (sample \#2) below 5 K. b) High temperature behavior (sample \#1) above 5 K. c) Plot of the amplitudes vs. temeperature in the range 30 to 50 T for both samples measured. The
solid line is a fit to a modified version of the standard Lifshitz-Kosevich formula for quantum oscillation amplitudes in metals (see text).

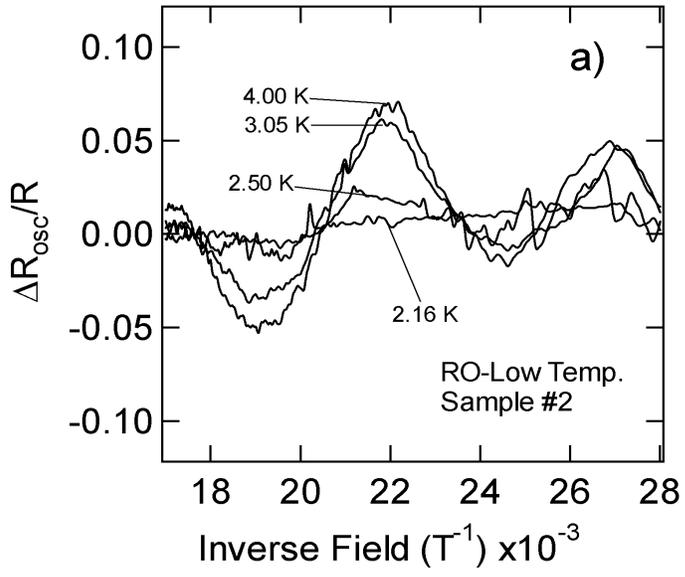
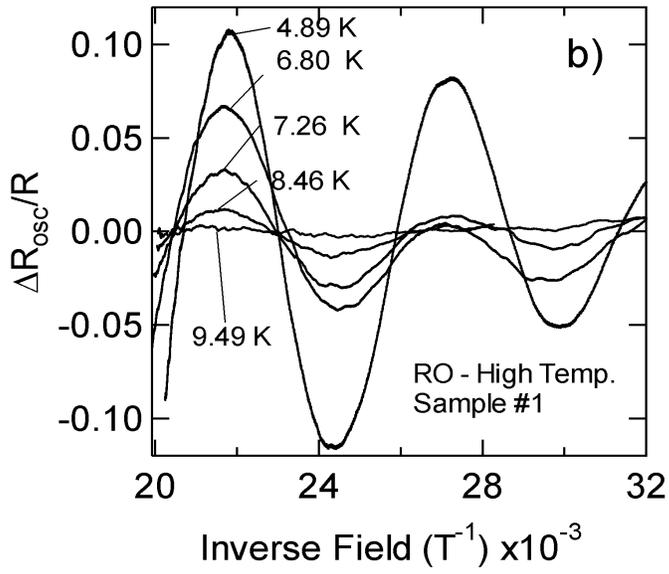
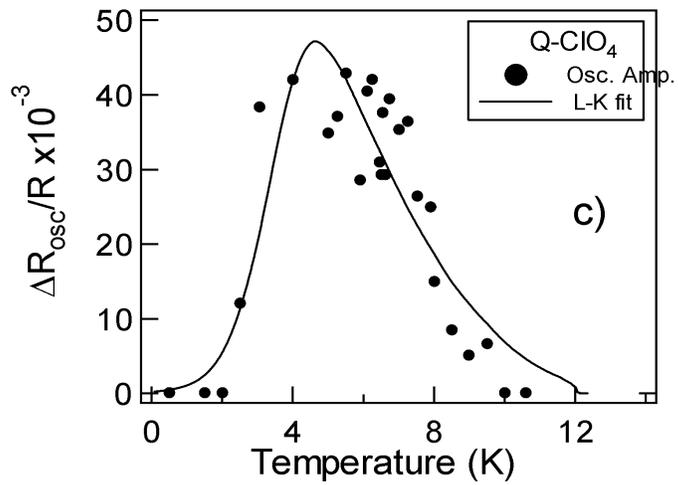
\end{document}